\documentclass[showpacs,preprintnumbers,amsmath,amssymb]{revtex4}

\usepackage{graphicx}
\usepackage{dcolumn}
\usepackage{bm}

\def\simge{\mathrel{%
       \rlap{\raise 0.511ex \hbox{$>$}}{\lower 0.511ex \hbox{$\sim$}}}}
\def\simle{\mathrel{
       \rlap{\raise 0.511ex \hbox{$<$}}{\lower 0.511ex \hbox{$\sim$}}}}

\begin{document}

\preprint{\sf BNL-NT-08/11\ \ 2008/April}

\title{Canonical partition function and finite density phase transition 
in lattice QCD}

\author{Shinji Ejiri}

\affiliation{Physics Department, Brookhaven National Laboratory,
Upton, New York 11973, USA}

\date{October 10, 2008}

\begin{abstract}
We discuss the nature of the phase transition for lattice QCD 
at finite temperature and density. 
We propose a method to calculate the canonical partition function 
by fixing the total quark number introducing approximations allowed 
in the low density region. 
An effective potential as a function of the quark number density 
is discussed from the canonical partition function. 
We analyze data obtained in a simulation of two-flavor QCD 
using p4-improved staggered quarks with bare quark mass $m/T = 0.4$ 
on a $16^3 \times 4$ lattice. 
The results suggest that the finite density phase transition 
at low temperature is of first order. 
\end{abstract}


\maketitle

\section{Introduction}
\label{sec:intro}

The study of the QCD phase diagram at nonzero temperature $(T)$ and 
quark chemical potential $(\mu_q)$ is one of the most important topics 
among studies of lattice QCD.
In particular, the study of the endpoint of the first order phase 
transition line in the $(T, \mu_q)$ plane is interesting 
both from the experimental and theoretical point of view. 
The existence of such a critical point is suggested by phenomenological 
studies \cite{AY,Bard,SRS}. 
The appearance of the critical point in the $(T, \mu_q)$ plane 
is closely related to hadronic fluctuations in heavy ion collisions 
and may be experimentally examined by an event-by-event analysis 
of heavy ion collisions.
Many trials have been made to find the critical point 
by first principle calculation in lattice QCD 
\cite{FK2,crtpt,BS03,BS05,KS06,dFP2,GG2,eji05,eji07}. 
However, no definite conclusion on this issue is obtained so far.

One of the interesting approaches is to construct the canonical 
partition function ${\cal Z}_{\rm C} (T,N)$ by fixing the total 
quark number $(N)$ or quark number density $(\rho)$ 
\cite{Gibbs86,Bar97,Mill87,Hase92,Alex05,Krat05}. 
The canonical partition function is obtained from the grand canonical 
partition function 
${\cal Z}_{\rm GC}(T,\mu_q)$ by an inverse Laplace transformation.
The relation between ${\cal Z}_{\rm GC}(T,\mu_q)$ and 
${\cal Z}_{\rm C} (T,N)$ is given by 
\begin{eqnarray}
{\cal Z}_{\rm GC}(T,\mu_q) 
= \int {\cal D}U \left( \det M(\mu_q/T) \right)^{N_{\rm f}} e^{-S_g}
= \sum_{N} \ {\cal Z}_{\rm C}(T,N) e^{N \mu_q/T}, 
\label{eq:cpartition} 
\end{eqnarray}
where $\det M$ is the quark determinant, $S_g$ is the gauge action, 
and $N_{\rm f}$ is the number of flavors.
$N_{\rm f}$ in this equation must be replaced by $N_{\rm f}/4$ 
when one uses a staggered type quark action. 
In order to investigate the net quark number giving the largest 
contribution to the grand canonical partition function at $(T, \mu_q)$, 
it is worth introducing an effective potential $V_{\rm eff}$ as 
a function of $N$, 
\begin{eqnarray}
V_{\rm eff}(N) 
\equiv - \ln {\cal Z}_{\rm C}(T,N) -N \frac{\mu_q}{T} 
= \frac{f(T,N)}{T} -N \frac{\mu_q}{T} ,
\label{eq:effp} 
\end{eqnarray}
where 
$f$ is the Helmholtz free energy. 
If there is a first order phase transition region, 
we expect this effective potential has minima at more than one value 
of $N$. At the minima, the derivative of $V_{\rm eff}$ satisfies 
\begin{eqnarray}
\frac{\partial V_{\rm eff}}{\partial N} (N, T, \mu_q ) 
= -\frac{\partial (\ln {\cal Z}_{\rm C})}{\partial N} (T,N) 
- \frac{\mu_q}{T} =0 .
\label{eq:dpotdn} 
\end{eqnarray}
Hence, in the first order transition region of $T$, we expect 
$\partial (\ln {\cal Z}_{\rm C})/ \partial N (T,N) \equiv - \mu_q^*/T$ 
takes the same value at different $N$. 
Here, $\mu_q^* (T,N)$ is the chemical potential which gives a minimum 
of the effective potential at $(T, N)$. 

\begin{figure}[t]
\begin{center}
\includegraphics[width=6.0in]{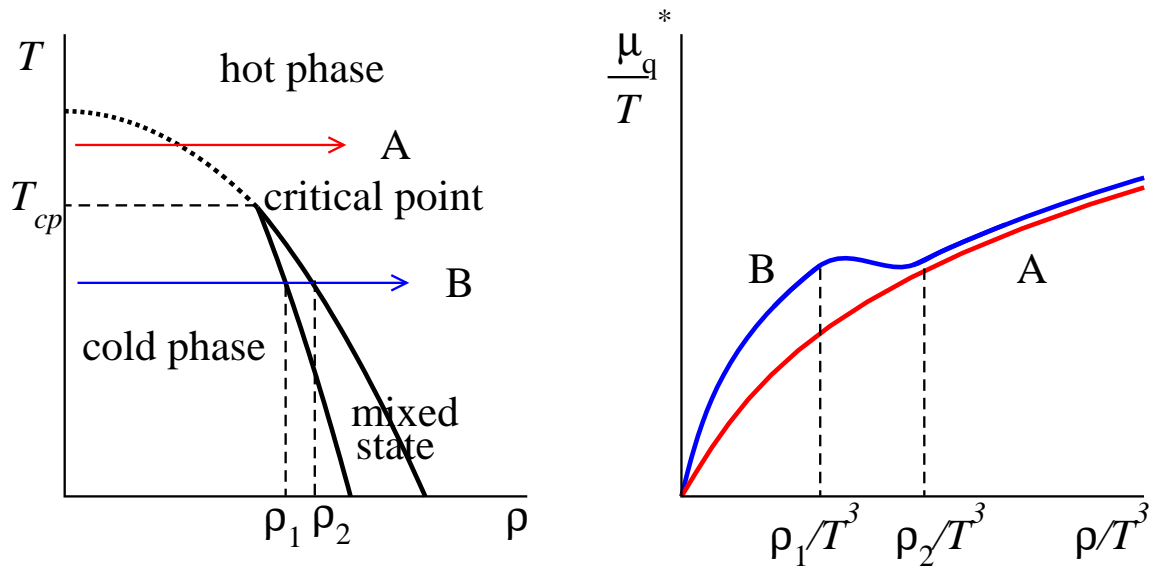}
\vskip -0.2cm
\caption{
Phase structure in the $(T, \rho)$ plane and the behavior of 
$\mu_q^*/T$ as a function of $\rho$.
}
\label{fig:phdiag}
\end{center}
\vskip -0.3cm
\end{figure}

The phase structure in the $(T, \rho)$ plane and the expected behavior 
of $\mu_q^*/T$ are sketched in the left and right panels of 
Fig.~\ref{fig:phdiag}, respectively. 
The thick lines in the left figure are the phase transition line. 
We expect that the transition is crossover at low density and becomes 
of first order at high density. 
Since two states coexist on the first order transition line, the phase 
transition line splits into two lines in the high density region, 
and the two states are mixed in the region between two lines. 
The expected behavior of $\mu_q^*$ along the lines A and B are shown 
in the right figure. 
When the temperature is higher than the temperature at the critical 
point $T_{pc}$ (line A), 
$\mu_q^*$ increases monotonically as the density increases. However, 
for the case below $T_{cp}$ (line B), this line crosses the mixed state. 
Because the two states of $\rho_1$ and $\rho_2$ are realized at 
the same time, $\mu_q^*$ does not increase in this region between 
$\rho_1$ and $\rho_2$. 

The Glasgow method \cite{Gibbs86,Bar97} has been a well-known method to 
compute the canonical partition function.
Recently, such a behavior at a first order phase transition has been 
observed by Kratochvila and de Forcrand in 4-flavor QCD with staggered 
fermions on a $6^3 \times 4$ lattice \cite{Krat05} calculating 
the quark determinant by the Glasgow algorithm.
However, with present day computer resources,
the study by the Glasgow method is difficult except on a small lattice. 
Therefore, it is important to consider a method available for 
a simulation on a large lattice. 
In this paper, we propose such a method for the calculation of 
the canonical partition function introducing approximations allowed 
in the low density region. 

The method proposed in this paper is based on the following ideas. 
We adopt a saddle point approximation for the inverse Laplace 
transformation from ${\cal Z}_{\rm GC}$ to ${\cal Z}_{\rm C}$. 
This approximation is valid when the volume size is sufficiently large. 
We moreover perform a Taylor expansion of $\ln \det M(\mu_q)$ 
in terms of $\mu_q$ around $\mu_q=0$ and calculate the expansion 
coefficients, as proposed in \cite{BS02}.
The Taylor expansion coefficients are rather easy to calculate 
by using the random noise method. 
The saddle point approximation is also based on a Taylor expansion 
around the saddle point. We estimate the Taylor expansion coefficients 
at the saddle point from the Taylor expansion at $\mu_q=0$. 
Although we must cut off this expansion at an appropriate order in $\mu_q$, 
this approximation is applicable when the saddle point is not far 
from $\mu_q=0$, and we can estimate the application range where 
the approximation is valid for each analysis. 

For the calculation of ${\cal Z}_{\rm C}$ two further technical 
problems must be solved. 
The first problem is the problem of importance sampling in Monte-Carlo 
simulations (overlap problem). Since the method proposed in this paper 
is a kind of reweighting method, the configurations which give important 
contributions will change when the weight factor is changed by $\mu_q$ 
\cite{Fodoreos,eji04}. 
To avoid this problem, we combine configurations generated at many 
simulation points $\beta =6/g^2$ covering a wide range of the temperature 
using a method introduced in \cite{Swen89}. 
The second problem is the sign problem. 
We must deal with an expectation value of a complex number in this method. 
If the fluctuation of the complex phase is large, the statistical error 
becomes larger than the mean value.
We use a technique introduced in \cite{eji07}. 
We consider the probability distribution function in terms of the complex 
phase of the complex operators.
We assume the distribution function is well approximated 
by a Gaussian function and perform the integration over the phase. 
Once this assumption is adopted, the sign problem is completely solved. 
This assumption is reasonable for sufficiently large volume and small 
$\mu_q/T$. 

During the process of this calculation, we will find out why the quark 
number density changes sharply at the transition point and why the 
density approaches the value of the free quark gas in the high density 
limit even at low temperature.
The configurations generated in $\mu_q=0$ simulations at low temperature 
are gradually suppressed as the density increases. 

In the next section, we explain the method to calculate the canonical 
partition function using the inverse Laplace transformation of 
${\cal Z}_{\rm GC}$ within a saddle point approximation. The problem 
of the Monte-Carlo sampling is discussed in Sec.~\ref{sec:analy}. 
The sign problem is discussed in Sec.~\ref{sec:phase}. 
We evaluate $\partial (\ln {\cal Z}_{\rm C})/ \partial N$  using data 
obtained with two-flavors of p4-improved staggered quarks in \cite{BS05}.
The result is shown in Sec.~\ref{sec:disc}. 
The behavior of $\partial (\ln {\cal Z}_{\rm C})/ \partial N$ suggests 
that the phase transition is of first order in the low temperature and 
high density region. 
Conclusions are given in Sec.~\ref{sec:conc}.

\section{Canonical partition function}
\label{sec:method}

We calculate the canonical partition function using $N_s^3 \times N_t$ 
lattice and investigate the effective potential $V_{\rm eff}(N)$. 
From Eq.~(\ref{eq:cpartition}), 
the canonical partition function can be obtained by an inverse Laplace 
transformation \cite{Mill87,Hase92,Alex05,Krat05},
\begin{eqnarray}
{\cal Z}_{\rm C}(T,N) = \frac{3}{2 \pi} \int_{-\pi/3}^{\pi/3} 
e^{-N (\mu_0/T+i\mu_I/T)} {\cal Z}_{\rm GC}(T, \mu_0+i\mu_I) \ 
d \left( \frac{\mu_I}{T} \right) ,
\label{eq:canonicalP} 
\end{eqnarray}
where $\mu_0$ is an appropriate real constant and $\mu_I$ is 
a real variable. Note that 
${\cal Z}_{\rm GC}(T, \mu_q +2\pi iT/3) = {\cal Z}_{\rm GC}(T, \mu_q)$
\cite{Rob86}.
The grand canonical partition function can be evaluated by 
the calculation of the following expectation value at $\mu_q=0$.
\begin{eqnarray}
\frac{{\cal Z}_{\rm GC}(T, \mu_q)}{{\cal Z}_{\rm GC}(T,0)}
= \frac{1}{{\cal Z}_{\rm GC}} \int {\cal D}U 
\left( \frac{\det M(\mu_q/T)}{\det M(0)} \right)^{N_{\rm f}}
(\det M(0))^{N_{\rm f}} e^{-S_g} 
= \left\langle 
\left( \frac{\det M(\mu_q/T)}{\det M(0)} \right)^{N_{\rm f}}
\right\rangle_{(T, \mu_q=0)} .
\label{eq:normZGC} 
\end{eqnarray}
However, with present day computer resources, the exact calculation 
is difficult except on small lattices. 
We consider an approximation which is valid for large volume and low density.
If we select a saddle point as $\mu_0$ in Eq.~(\ref{eq:canonicalP}) 
when the volume is sufficiently large, 
the information which is needed for the integral is only the value of $\det M$ 
around the saddle point. 
Furthermore, if we restrict ourselves to study the low density region, 
the value of $\det M (\mu_q/T)$ near the saddle point can be estimated 
by a Taylor expansion around $\mu_q=0$. 
The calculations by the Taylor expansion are much cheaper than the exact calculations and the studies 
using large lattices are possible. 

First, we perform the integral in Eq.~(\ref{eq:canonicalP}) by 
a saddle point approximation.
We denote the quark number density in a lattice unit and physical unit as
$\bar{\rho}=N/N_s^3$ and $\rho/T^3=\bar{\rho} N_t^3$, respectively. 
We assume that a saddle point $z_0$ exists in the complex $\mu_q/T$ 
plane for each configuration, which satisfies
\begin{eqnarray}
\left[ D'(z) - \bar{\rho} \right]_{z=z_0} =0, 
\label{eq:saddle}
\end{eqnarray}
where $(\det M(z)/\det M(0))^{N_{\rm f}} = \exp[N_s^3 D(z)] $ and 
$D'(z)=dD(z)/dz$.

We then perform a Taylor expansion around the saddle point and obtain 
the canonical partition function, 
\begin{eqnarray}
{\cal Z}_{\rm C}(T, \bar{\rho} V) &=& \frac{3}{2 \pi} {\cal Z}_{\rm GC}(T,0) 
\left\langle \int_{-\pi/3}^{\pi/3} 
e^{-N (z_0+ix)} \left( \frac{\det M (z_0+ix)}{\det M(0)} \right)^{N_{\rm f}}
dx \right\rangle_{(T, \mu_q=0)} \nonumber \\
&=& \frac{3}{2 \pi} {\cal Z}_{\rm GC}(T,0) \left\langle \int_{-\pi/3}^{\pi/3} 
\exp \left[ V \left(D (z_0) - \bar{\rho} z_0 
- \frac{1}{2} D'' (z_0) x^2 + \cdots \right) \right] 
dx \right\rangle_{(T, \mu_q=0)} \nonumber \\
& \approx & \frac{3}{\sqrt{2 \pi}} {\cal Z}_{\rm GC} (T,0)
\left\langle \exp \left[ V \left( D(z_0) - \bar{\rho} z_0 \right) \right] 
e^{-i \alpha/2} \sqrt{ \frac{1}{V |D''(z_0)|}}
\right\rangle_{(T, \mu_q=0)} .
\label{eq:zcspa}
\end{eqnarray}
Here, $ D''(z) = d^2 D(z) /dz^2,$ $V \equiv N_s^3$ and 
$D''(z)=|D''(z)| e^{i \alpha}$.
We chose a path which passes the saddle point.
Higher order terms in the expansion of $D(z)$ becomes negligible when 
the volume $V$ is sufficiently large, since the saddle point approximation 
is a $1/V$ expansion. 

Next, we calculate the quark determinant by the Taylor expansion 
around $\mu_q=0$. 
We define 
\begin{eqnarray}
D_n = \frac{1}{n! N_s^3 N_t} \left[ 
\frac{\partial^n \ln \det M(\mu_q/T)}{ \partial (\mu_q/T)^n} 
\right]_{\mu_q=0}.
\end{eqnarray}
The even derivatives of $\ln \det M$ 
are real and the odd derivatives are purely imaginary \cite{BS02}. 
The calculation of $D_n$ is rather easy using the stochastic noise method. 
$D(z), D'(z)$ and $D''(z)$ in Eq.~(\ref{eq:saddle}) and (\ref{eq:zcspa}) 
can be evaluated by 
\begin{eqnarray}
D (z)= N_{\rm f} N_t \sum_{n=1}^{\infty} D_n z^n , \hspace{3mm}
D'(z)= N_{\rm f} N_t \sum_{n=1}^{\infty} n D_n z^{n-1} , \hspace{3mm}
D''(z) = N_{\rm f} N_t \sum_{n=2}^{\infty} n(n-1) D_n z^{n-2} . 
\end{eqnarray}
Because ${\rm Im}(D_1) \ll {\rm Re}(D_2)$, the saddle point, 
i.e. the solution of  Eq.~(\ref{eq:saddle}), is distributed near the real 
axis and ${\rm Re} (z_0)$ increases as $\rho$ increases. 
Moreover, for the case that the saddle point $z_0$ is on the real axis, 
the saddle point condition is the same as the condition where 
${\rm Re}(D(z)- \bar{\rho} z)$ is minimized. 
Hence, $\exp[V{\rm Re}(D(z_0) - \bar{\rho} z_0)]$ in Eq.~(\ref{eq:zcspa}) 
decreases exponentially as ${\rm Re} (z_0)$ increases.

In this study, we want to focus on the derivative of the effective potential 
with respect to $N$ or $\rho$.
Since the effective potential $V_{\rm eff}(N)$ is minimized 
in the thermodynamic limit, i.e. 
$\partial \log {\cal Z}_{\rm C}/ \partial N + \mu_q/T=0$, 
we denote the derivative by
\begin{eqnarray}
\frac{\mu_q^*}{T} 
= - \frac{\partial \ln {\cal Z}_{\rm C} (T,N) }{\partial N}
\ = \ - \frac{1}{V} \frac{\partial \ln {\cal Z}_{\rm C} (T, \bar{\rho} V)}
{\partial \bar{\rho}} .
\label{eq:chem}
\end{eqnarray}
Within the framework of the saddle point approximation, 
this quantity can be evaluated by 
\begin{eqnarray}
\frac{\mu_q^*}{T} 
\approx \frac{
\left\langle z_0 \ \exp \left[ V \left( D(z_0)
- \bar{\rho} z_0 \right) \right] 
e^{-i \alpha /2} \sqrt{ \frac{1}{V |D''(z_0)|}}
\right\rangle_{(T, \mu_q=0)}}{
\left\langle \exp \left[ V \left( D(z_0) 
- \bar{\rho} z_0 \right) \right] 
e^{-i \alpha /2} \sqrt{ \frac{1}{V |D''(z_0)|}}
\right\rangle_{(T, \mu_q=0)}}. 
\label{eq:chemap}
\end{eqnarray}
This equation is similar to the formula of the reweighting method 
for finite $\mu_q$. 
The operator in the denominator corresponds to a reweighting factor, 
and $\mu_q^* /T$ is an expectation value of the saddle 
point calculated with this modification factor.
We denote the real and imaginary parts of the logarithm of the weight 
factor by $F$  and $\theta$,
\begin{eqnarray}
F + i \theta \equiv V \left( N_{\rm f} 
N_t \sum_{n=1}^{\infty} D_n z_0^n - \bar{\rho} z_0 \right) 
-\frac{1}{2} \ln [V |D''(z_0)|] -\frac{i \alpha}{2} .
\label{eq:logw}
\end{eqnarray}
We define the complex phase of the weight factor by $\theta$ 
and the absolute value of the reweighting factor is $\exp(F)$.
This weight factor plays an important role around the phase transition point 
at finite density. 

In the calculation to derive Eq.~(\ref{eq:zcspa}), we replaced the order of 
the path integral of gauge fields and the integral for the inverse 
Laplace transformation. This replacement is essentially important. 
If one calculates ${\cal Z}_{\rm C}$ from ${\cal Z}_{\rm GC}$ 
using Eq.~(\ref{eq:canonicalP}) after the path integral, 
the equation which is satisfied at a saddle point is 
\begin{eqnarray}
N= \bar{\rho}V = \partial (\ln {\cal Z}_{\rm GC})/ \partial (\mu_q/T). 
\end{eqnarray}
Hence, in the thermodynamic limit, i.e. when we ignore the finite volume 
correction, $\mu_q^*$ is just equal to the inverse function of 
$\rho (\mu_q)$ at the saddle point.
Therefore, $\rho (\mu_q)$ must be a discontinuous function or a multivalued 
function at a first order phase transition to obtain the behavior of 
$\mu_q^* (\rho)$ shown in Fig.~\ref{fig:phdiag}.
However, if we calculate $\ln {\cal Z}_{\rm GC}$ by a Taylor expansion 
in $\mu_q$ at a temperature in the hadron phase, $\rho (\mu_q)$ 
cannot be a discontinuous function. 

In this study, we use Eq.~(\ref{eq:chemap}). 
As we will discuss in detail, we can obtain the behavior of $\mu_q^*/T$ 
suggesting a first order phase transition, although the calculations of 
$z_0$ and the weight factor in Eq.~(\ref{eq:chemap}) are based on the 
Taylor expansion. The important point is that the weight factor 
$\exp(F+i \theta)$ gives the same effect when the temperature changed 
and configurations which give important contributions to the calculation 
of $\mu_q^*/T$ change gradually as $\rho$ increases. 
Hence $\mu_q^*/T$ does not need to increase monotonously as a function 
of $\rho$ even if the saddle point $z_0$ is a monotonous function of 
$\rho$ for each configuration.

\section{Monte-Carlo analysis and reweighting method for $\beta$-direction}
\label{sec:analy}

\begin{figure}[t]
\begin{center}
\includegraphics[width=3.1in]{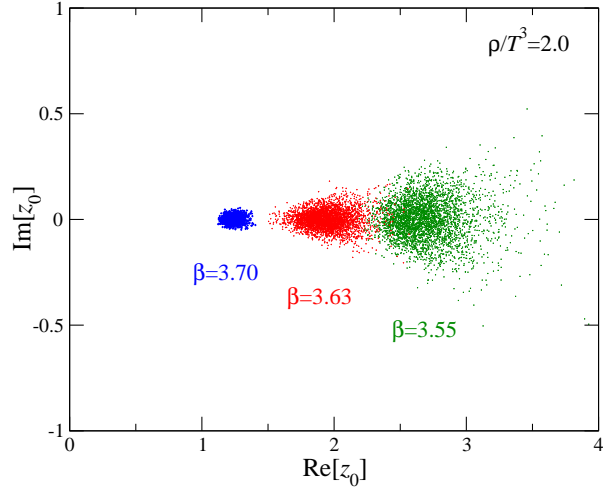}
\vskip -0.2cm
\caption{
Distribution of the saddle point at $\beta=3.55, 3.63, 3.70$ 
with $\rho/T^3=2.0$.
}
\label{fig:z0dis}
\end{center}
\vskip -0.3cm
\end{figure} 

\begin{figure}[t]
\begin{center}
\includegraphics[width=3.1in]{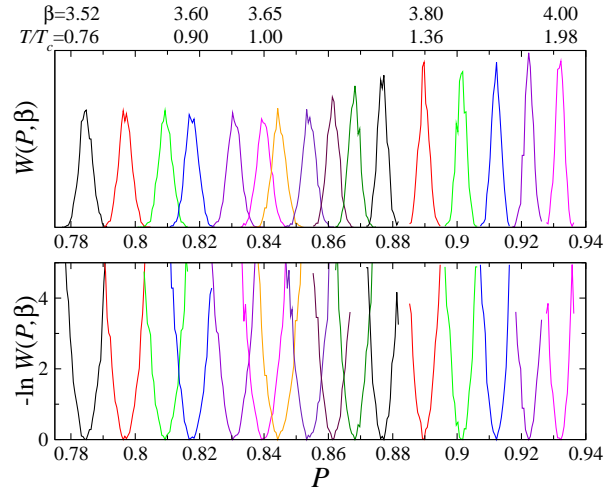}
\vskip -0.2cm
\caption{
Plaquette histogram $w(P, \beta)$ at $\mu_q=0$.
}
\label{fig:plhis}
\end{center}
\vskip -0.3cm
\end{figure} 

\begin{figure}[t]
\begin{center}
\includegraphics[width=3.1in]{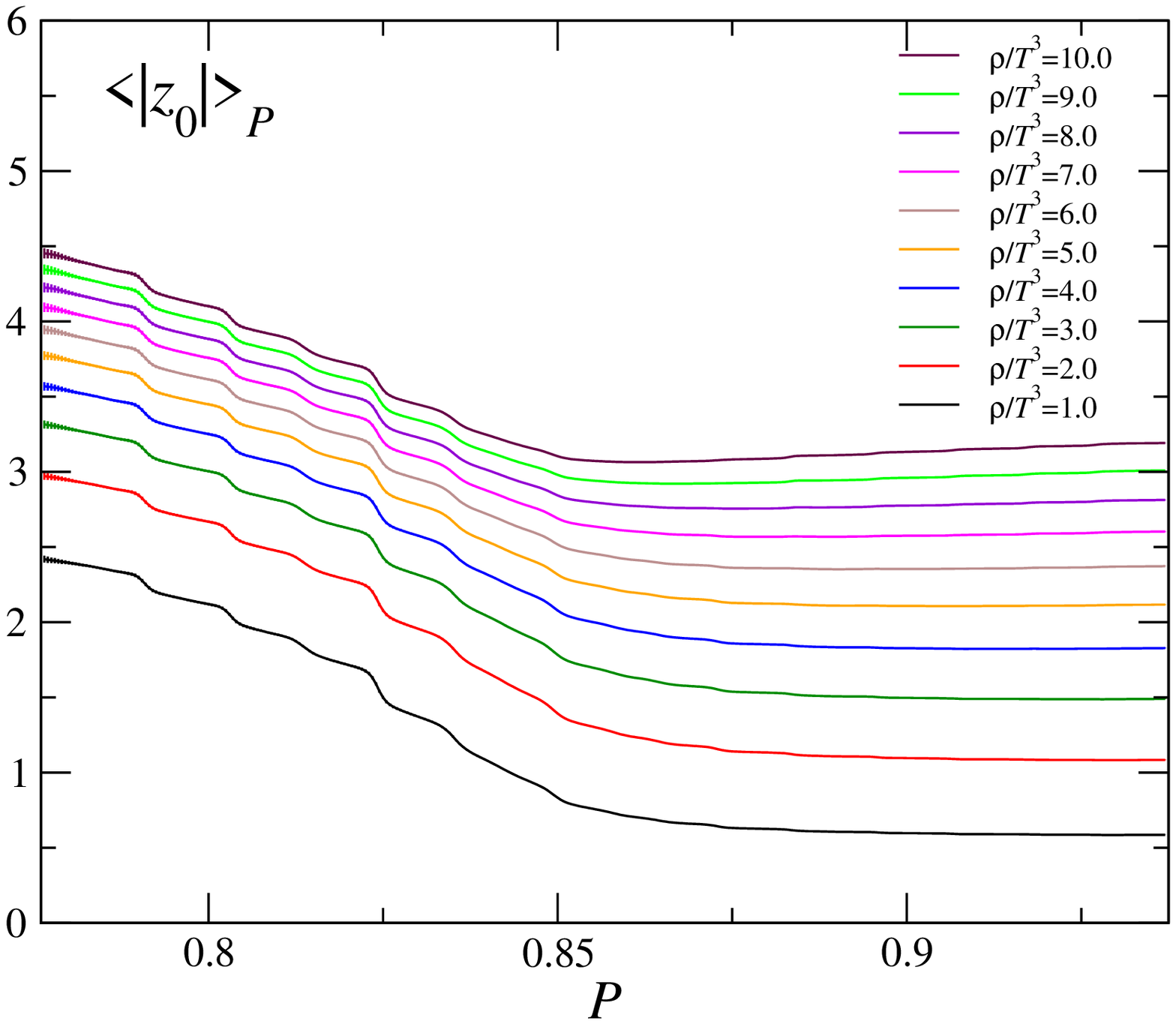}
\vskip -0.2cm
\caption{
Expectation value of $|z_0|$ with fixed $P$ for each $\rho/T^3$. 
}
\label{fig:rahis}
\end{center}
\vskip -0.3cm
\end{figure} 

\begin{figure}[t]
\begin{center}
\includegraphics[width=3.1in]{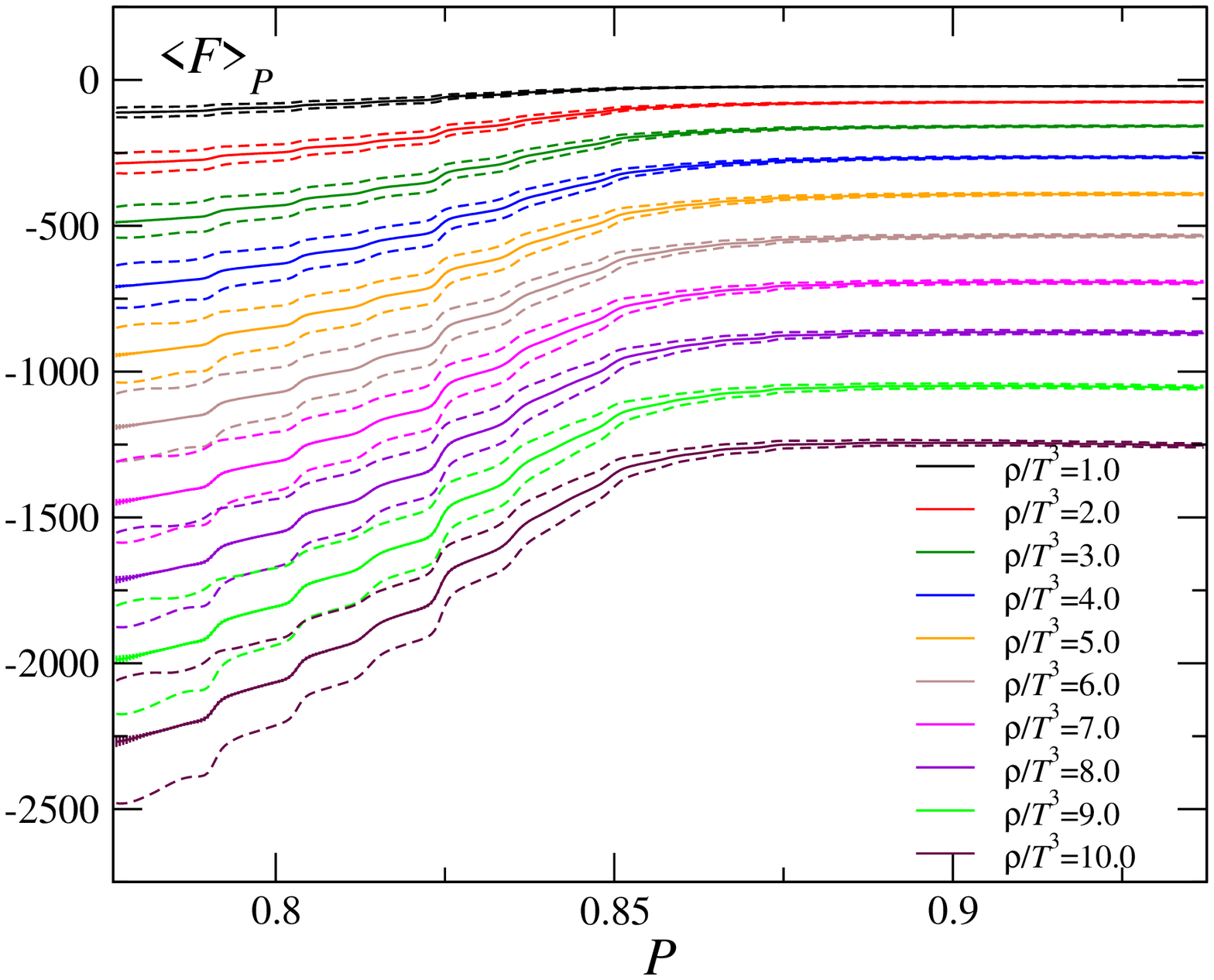}
\vskip -0.2cm
\caption{
Expectation vale of $F$ with fixed $P$ for each $\rho/T^3$. 
}
\label{fig:ffhis}
\end{center}
\vskip -0.3cm
\end{figure} 

\begin{figure}[t]
\begin{center}
\includegraphics[width=3.1in]{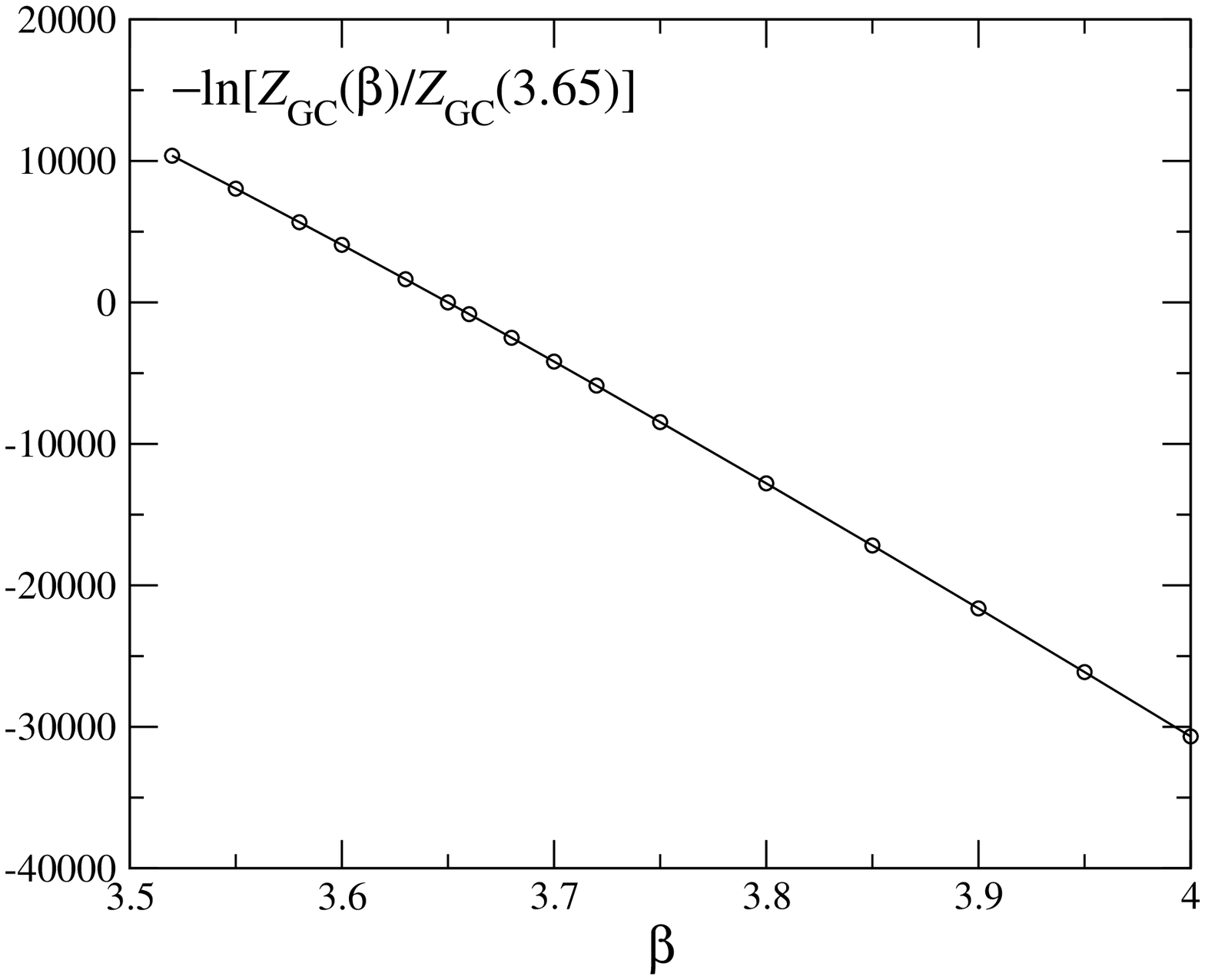}
\vskip -0.2cm
\caption{
$\beta$ dependence of $- \ln {\cal Z}_{\rm GC}$ determined by 
the consistency condition Eq.~(\ref{eq:mprcc}).
}
\label{fig:lnz}
\end{center}
\vskip -0.3cm
\end{figure}

We analyze data obtained in a simulation of two-flavor QCD using 
p4-improved staggered quarks with bare quark mass $m/T = 0.4$ \cite{BS05}. 
These data are obtained at 16 simulation points from $\beta=3.52$ 
to $4.00$.
The corresponding temperature normalized by the pseudocritical 
temperature is in the range of $T/T_c= 0.76$ to $1.98$, 
and the pseudocritical point $(T/T_c=1)$ is $\beta_{pc} \approx 3.65$. 
The ratio of pseudoscalar and vector meson masses is 
$m_{\rm PS}/m_{\rm V} \approx 0.7$ at $\beta=3.65$.
The lattice size $N_{\rm site}=N_s^3 \times N_t$ is $16^3 \times 4$. 
The number of configurations $N_{\rm conf.}$ is 1000 -- 4000 for each $\beta$. 
Further details on the simulation parameters are given in \cite{BS05}.

We use the data of the Taylor expansion coefficients $D_n$ up to $O(\mu_q^6)$.
The saddle point is found by the following procedure. 
(1) Because absolute values of the odd terms of $D_n$ are much smaller than 
the even terms, we first find a solution when the odd terms are neglected, 
i.e. a solution of $N_{\rm f} N_t \sum_n 2nD_{2n} z^{2n-1} - \bar{\rho}=0$. 
The odd terms of $D_n$ are purely imaginary and the even terms are real. 
Although some fake solutions may appear at large $z$ due to the truncation 
of the higher order terms, the solution is found on the real axis 
in the low density regime. 
(2) Next, in the vicinity of this solution, we calculate 
$r^2 \equiv |N_{\rm f} N_t \sum_n nD_{n} z^{n-1} - \bar{\rho}|^2$ 
and find the point where $r^2$ is zero. 
For the data we used in this analysis, the saddle point could be found 
for every configuration by this procedure except in the very low density 
region of $\rho/T^3 < 0.37$. 
Figure \ref{fig:z0dis} shows an example of the distribution of the saddle 
points, obtained at $\beta=3.55, 3.63, 3.70$ with $\rho/T^3=2.0$. 

For the calculation of the derivative of $\ln {\cal Z}_{\rm C}$, 
the application of the reweighting method for $\beta$-direction is crucial. 
Configurations in a Monte-Carlo simulation are generated with 
the probability in proportion to the product of the weight factor 
$(\det M)^{N_{\rm f}} e^{-S_g}$ 
and the state density of the link fields $\{ U_{\mu}(x) \}$. 
The expectation value is then estimated by taking an average 
of the operator ${\cal O}[U_{\mu}]$ over the generated configurations 
$\{ U_{\mu}(x) \}$.
\begin{eqnarray}
\langle {\cal O} \rangle_{(\beta)} \approx 
\frac{1}{N_{\rm conf.}} \sum_{ \{ U_{\mu}(x) \} } {\cal O}[U_{\mu}].
\end{eqnarray}
However, if the value of ${\cal O}[U_{\mu}]$ changes very much during a 
Monte-Carlo simulation and the change of ${\cal O}[U_{\mu}]$ is much 
larger than the size of the probability $p({\cal O}, \beta)$, 
the Monte-Carlo method is no longer valid. 
For example, configurations which have large 
${\cal O} \times p({\cal O}, \beta)$ 
are important for the evaluation of $\langle {\cal O} \rangle_{(\beta)}$, 
but such configurations are not generated if $p({\cal O}, \beta)$ is too small. 
Such a problem occurs in the calculation of Eq.~(\ref{eq:chemap}). 
This problem is called an ``overlap problem.''

To clarify the problem of the importance sampling, 
we rewrite Eq.~(\ref{eq:chemap}) as
\begin{eqnarray}
\frac{\mu_q^*}{T}
= \frac{ \int \langle z_0 \exp[F+i \theta] \rangle_P w(P, \beta) dP
}{ \int \langle \exp[F+i \theta] \rangle_P w(P, \beta) dP } .
\label{eq:sdm}
\end{eqnarray}
$P$ is the plaquette value, 
$\langle \cdots \rangle_P$ denotes the expectation value for fixed $P$ at 
$\mu_q=0$, and $w(P, \beta)$ is the probability distribution of $P$ at $\beta$,
\begin{eqnarray}
w(P', \beta) = \int {\cal D}U \delta (P- P') (\det M(0))^{N_{\rm f}} e^{-S_g(\beta)}
\propto \langle \delta (P- P') \rangle_{(\beta, \mu_q=0)} .
\label{eq:pdis}
\end{eqnarray}
This kind of analysis is called the density of state method 
\cite{Swen88,Gock88,Aloi00,JN01,FKS07,Ta04}.
We define the average plaquette as $P=-S_g/(6N_{\rm site} \beta)$ 
for later discussions. 
This $P$ is the plaquette value for the standard gauge action, 
but is a linear combination of Wilson loops for improved gauge actions. 
Because
\begin{eqnarray}
\langle X \rangle_{P'} \equiv 
\frac{\langle X \delta (P- P') \rangle_{(\beta, \mu_q=0)}}{
\langle \delta (P- P') \rangle_{(\beta, \mu_q=0)}}
= \frac{\int {\cal D}U X \delta (P- P') (\det M(0))^{N_{\rm f}}}
{\int {\cal D}U \delta (P- P') (\det M(0))^{N_{\rm f}}},
\end{eqnarray}
$\langle X \rangle_{P}$ is independent of $\beta$ for an operator 
$X$ which does not depend on $\beta$ explicitly.
Hence, $\langle X \rangle_{P}$ can be computed at an appropriate $\beta$.
The probability distribution functions $w(P)$ and $-\ln w(P)$ are 
given in Fig.~\ref{fig:plhis}. 
We show the values of $\beta$ and corresponding $T/T_c$ above these figures. 
To obtain $w(P)$, 
we grouped the configurations by the value of $P$ into blocks 
and counted the number of configurations in these blocks. 
$-\ln w(P)$ is normalized by the minimum value for each $\beta$.
Because the transition from the hadron phase to the quark-gluon phase is a crossover for two-flavor QCD, the distribution is always of Gaussian type, 
and the width of the distribution becomes narrower as the volume increases.
Moreover, since the suppression factor is $\exp(6N_{\rm site} \beta P)$, 
the peak position of the distribution $w(P)$ moves to the right 
as $\beta$ increases.

We also calculate the expectation value of $|z_0|$ and $F$ when $P$ is fixed, 
\begin{eqnarray}
\langle |z_0| \rangle_{P'} = \left. \langle |z_0| \delta (P- P') \rangle 
\right/ \langle \delta (P- P') \rangle , \hspace{5mm}
\langle F \rangle_{P'} = \left. \langle F \delta (P- P') \rangle 
\right/ \langle \delta (P- P') \rangle .
\end{eqnarray}
The result of $\langle |z_0| \rangle_{P}$ is plotted in Fig.~\ref{fig:rahis}, 
and solid lines in Fig.~\ref{fig:ffhis} are $\langle F \rangle_{P}$ 
for each $\rho/T^3$. 
(Dashed lines will be explained in the next section.)
For the calculation of these quantities, 
we use the delta function approximated by a Gaussian function,
$\delta(x) \approx 1/(\Delta \sqrt{\pi}) \exp[-(x/\Delta)^2]$. 
If making $\Delta$ small, this approximation becomes better but 
statistical errors become larger. 
Hence, the size of $\Delta$ must be adjusted appropriately. 
In this study, we adopt $\Delta=0.0025$.

Let us now consider $\langle \exp[F+i \theta] \rangle_P \times w(P)$ 
in Eq.~(\ref{eq:sdm}). 
Since $\langle F \rangle_P$ increases linearly for $P \simle 0.85$, 
Fig.~\ref{fig:ffhis} suggests that $\langle \exp[F+i \theta] \rangle_P$ 
increases exponentially as $P$ increases in this range. 
Moreover, the slope of $\langle F \rangle_P$ is increasing 
as $\rho/T^3$ increases. 
Therefore, for large $\rho/T^3$, this 
$\langle \exp[F+i \theta] \rangle_P \times w(P)$ may not decrease 
even if $w(P)$ decreases exponentially at the tail of the distribution 
generated by a simulation. 
If the value of $\langle \exp[F+i \theta] \rangle_P \times w(P)$ is 
still large even in the region where the configurations are not generated, 
the calculation by the Monte-Carlo method is completely wrong. 
Because the width of $w(P)$ becomes narrow for large $V$, 
it is essentially important to solve this problem if we want to use 
a large lattice which is required for the saddle point approximation. 

To avoid this problem, we combine all configurations obtained in 
simulations with many different $\beta$ values, using the Ferrenberg 
and Swendsen method \cite{Swen89}.
The expectation value $\left\langle {\cal O} \right\rangle_{\beta}$ can be 
calculated from the data obtained by more than one simulation point, 
$\beta_i$ $(i=1,2, \cdots, N_{\beta})$ by the following equation: 
\begin{eqnarray}
\left\langle {\cal O} \right\rangle_{\beta}
\approx 
\frac{ \left\langle {\cal O} G(\beta,P) \right\rangle_{\rm all}
}{ \left\langle G(\beta,P) \right\rangle_{\rm all}}.
\label{eq:evmultb}
\end{eqnarray}
Here, the weight factor $G(\beta, P)$ is
\begin{eqnarray}
G(\beta,P)=\frac{e^{6 N_{\rm site} \beta P}}{
\sum_{i=1}^{N_{\beta}} N_i e^{6 N_{\rm site} \beta_i P} 
{\cal Z}_{\rm GC}^{-1} (\beta_i)} ,
\label{eq:betarew}
\end{eqnarray}
where $N_i$ is the number of configurations at simulation points $\beta_i$ 
and $\left\langle \cdots \right\rangle_{\rm all}$ means the average over 
all configurations generated at all $\beta_i$.
The derivation of this equation is given in Appendix A.

The partition function ${\cal Z}_{\rm GC}(\beta_i)$ is 
determined by a consistency condition for each $i$, 
\begin{eqnarray}
{\cal Z}_{\rm GC}(\beta_i) 
= \left\langle G(\beta_i ,P) \right\rangle_{\rm all}
\label{eq:mprcc}
\end{eqnarray}
This equation can be solved except for the normalization factor. 
The result of $-\ln[{\cal Z}_{\rm GC}(\beta)/ {\cal Z}_{\rm GC}(3.65)]$ 
is plotted in Fig.~\ref{fig:lnz}. 

We should note that $G(\beta, P)$ is independent of the simulation 
points $\beta_i$ at which the operators are measured and the expectation 
value is simply given by the average over all configurations 
generated at many $\beta$. 
If we perform simulations at many different $\beta$ and combine 
the data, the configurations are distributed in a wide range of 
$P$. (See Fig.~\ref{fig:plhis}.) 
Among these configurations, important configurations for each 
calculation are selected by ${\cal O}$ and $G(\beta,P)$ automatically. 
This method is particularly important when the volume is large, since 
the distribution $w(P)$ is narrow if we generate configurations on 
a large lattice with single $\beta$. 
The overlap problem is solved by this method. However, the statistical 
error is enlarged by the fluctuations of $\exp[F+i \theta]$ when 
the density is increased, hence the application range of $\rho$ is 
determined by the statistical error. 
Also, we should check that the important configurations are within 
the range of the plaquette distribution for each calculation. 
We will discuss this point in Sec.~\ref{sec:disc} again.

\section{Sign problem}
\label{sec:phase}

\begin{figure}[t]
\begin{center}
\includegraphics[width=2.0in]{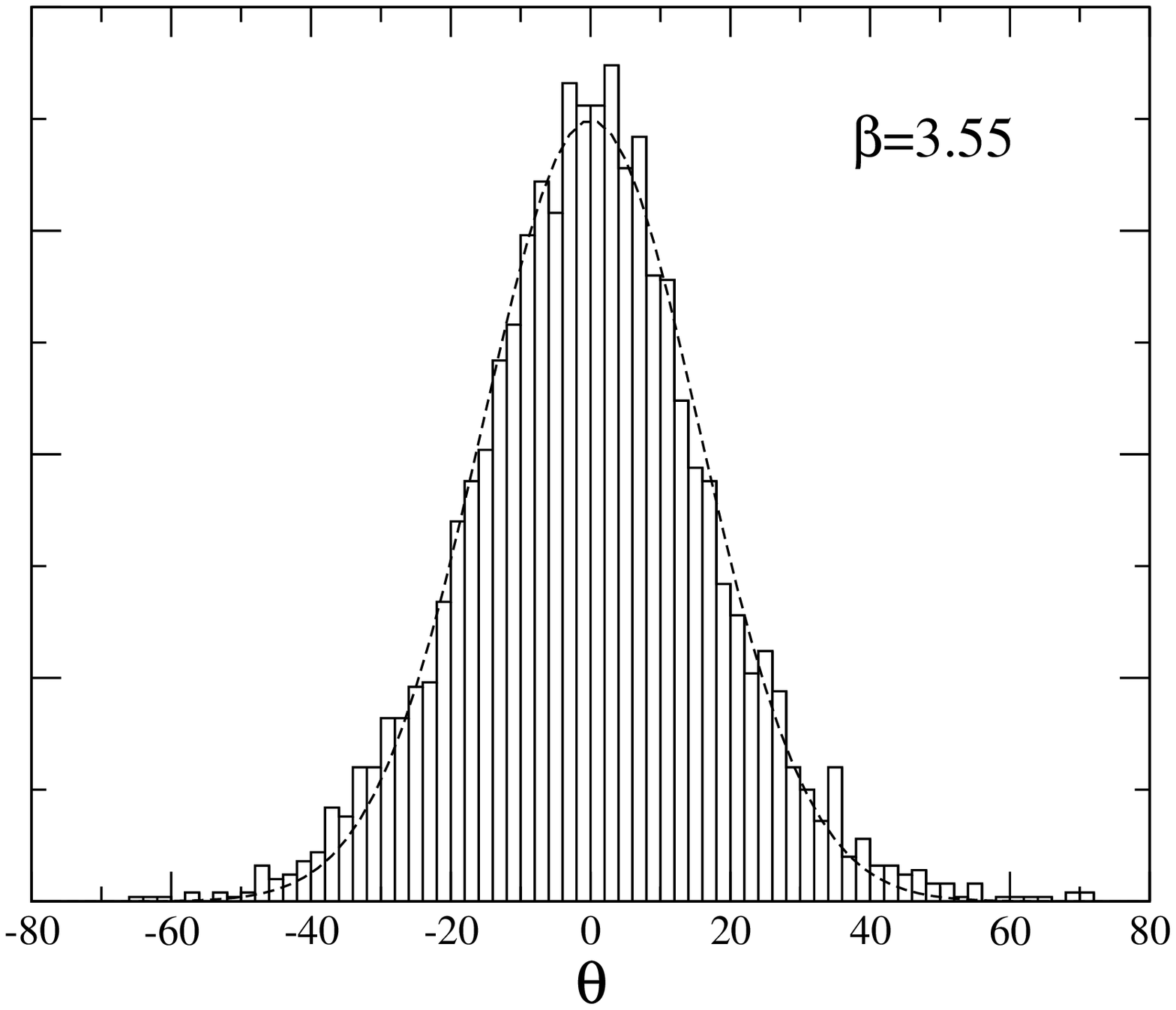}
\hskip 0.5cm
\includegraphics[width=2.0in]{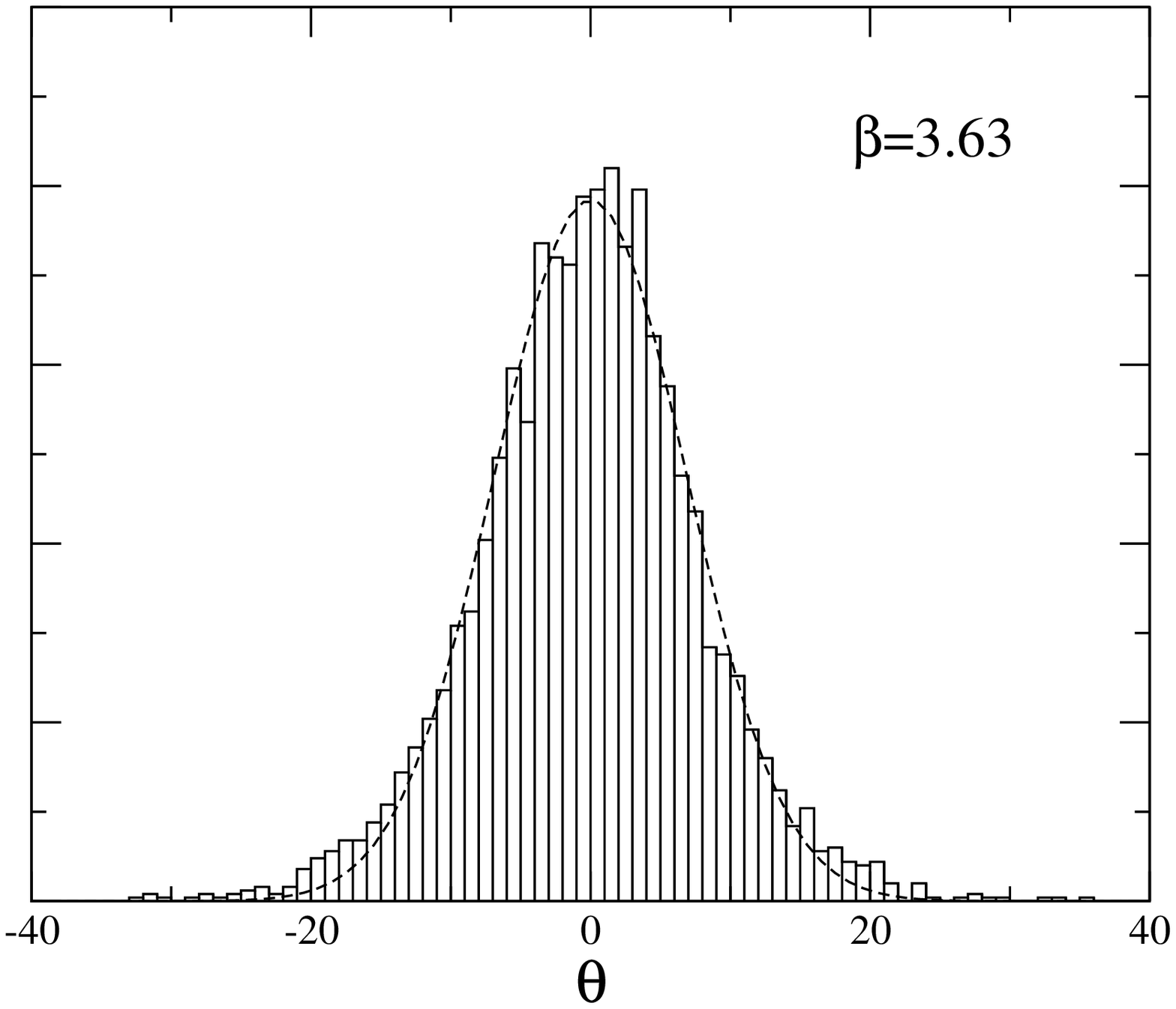}
\hskip 0.5cm
\includegraphics[width=2.0in]{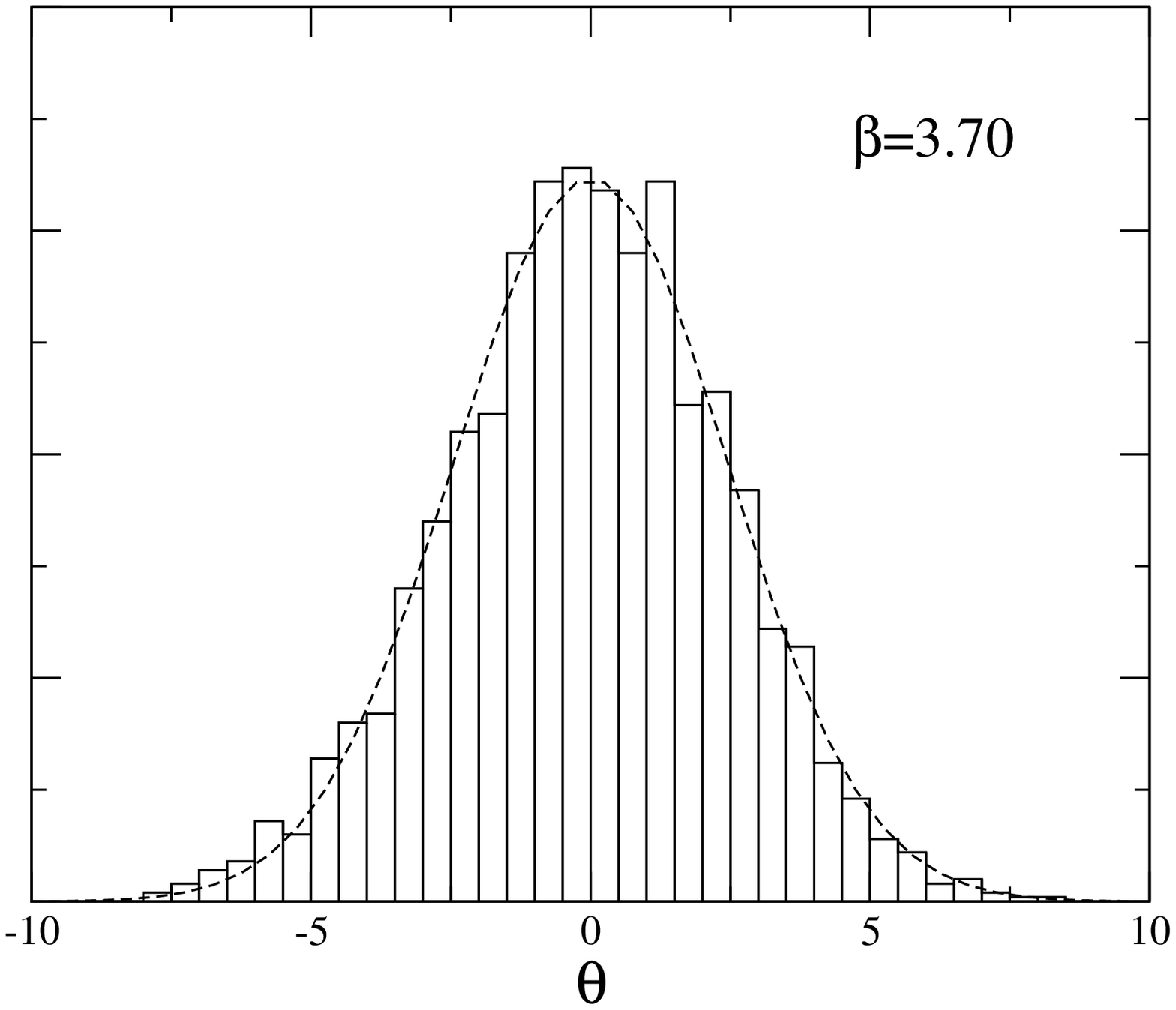}
\vskip -0.2cm
\caption{Histograms of the complex phase $\theta$
for $\rho/T^3=2.0$ 
at $\beta=3.55$ (left), $3.63$ (middle) and $3.70$ (right).
Dashed lines are the fit results by Gaussian functions.}
\label{fig:phhist}
\end{center}
\vskip -0.3cm
\end{figure}

Next, we discuss the sign problem 
that also shows up in the calculation of the canonical partition function. 
Because $z_0$ and the additional weight factor in Eq.~(\ref{eq:chemap})
are complex numbers, the calculation of Eq.~(\ref{eq:chemap}) suffers from
the sign problem \cite{eji04,spli06}. 
If the weight factor changes the sign frequently, both the numerator 
and denominator of Eq.~(\ref{eq:chemap}) become smaller than 
their statistical errors.
To avoid the sign problem, we use a method proposed in \cite{eji07}.
As in Eq.~(\ref{eq:logw}), 
we define the complex phase of the weight factor by
\begin{eqnarray}
\theta \equiv {\rm Im} \left[ V \left( N_{\rm f} 
N_t \sum_{n=1}^{\infty} D_n z_0^n - \bar{\rho} z_0 \right) \right]
-\frac{\alpha}{2} .
\label{eq:theta}
\end{eqnarray}
In this definition, $\theta$ is not restricted to 
the range from $-\pi$ to $\pi$ because 
there is no reason that the imaginary part of Eq.~(\ref{eq:logw}) must 
be in the finite range. 
In fact, this quantity becomes larger as the volume increases. 

It has been discussed in \cite{eji07} that histograms of $D_n$ 
are well approximated by Gaussian functions if a simulation is performed at 
a point away from the critical point with sufficiently large volume. 
The Taylor expansion coefficients in Eq.~(\ref{eq:logw}) are 
given by combinations of traces of products of 
$\partial^n M/ \partial (\mu_q/T)^n$ and $M^{-1}.$ 
(See the appendix of \cite{BS05}.) 
Therefore, $D_n$ are obtained by the sum 
of the diagonal elements of such matrices. 
When the correlation among the diagonal elements is small and 
the volume is sufficiently large, the distribution functions of 
the expansion coefficients and $\theta$ should be of Gaussian type 
due to the central limit theorem. 
For example, the diagonal element of the first coefficient, 
${\rm Im} [\partial (\ln \det M)/ \partial (\mu_q/T)]= 
{\rm Im} [{\rm tr} [M^{-1} (\partial M/ \partial (\mu_q/T))]]$, is 
the imaginary part of the local number density operator at $\mu_q=0$. 
If the spatial density correlation is not very strong, 
a Gaussian distribution is expected. 

Figure \ref{fig:phhist} is the histogram of the complex phase 
$\theta$ at $\rho/T^3=2.0$ for the two-flavor QCD simulations with 
p4-improved staggered quarks
at $\beta=3.55$ (left), $3.63$ (middle), and $3.70$ (right). 
These figures suggest that the distribution of $\theta$ is 
well approximated by a Gaussian function. We fit the data of 
the histogram by a Gaussian function. 
Dashed lines are the fit results. 
The width of the Gaussian function is different for each distribution 
obtained by a different parameter. 
If we restrict the phase to the range from $-\pi$ to $\pi$ by 
subtracting $2 \pi n$ ($n$: integer), the complex phase distribution 
is almost flat for the case that the width is much larger than $\pi$, 
and the flatness indicates the seriousness of the sign problem. 
However, the measurement of the width of the Gaussian distribution 
is easier than the estimation of the flatness of the restricted 
phase distribution. 

Once we assume a Gaussian distribution for $\theta$, 
the problem of complex weights can be avoided.
We introduce the probability distribution $\bar{w}$ as 
a function of the plaquette $P$, $F$, and $\theta$, 
\begin{eqnarray}
\bar{w}(P', F', \theta') \equiv 
\int {\cal D}U \delta (P'-P) \delta (F'-F)
\delta (\theta' - \theta) 
(\det M(0))^{N_{\rm f}} e^{6\beta N_{\rm site} P}.
\end{eqnarray}
The distribution function itself is defined as an expectation value 
at $\mu_q=0$, 
however $F$ and $\theta$ are functions of $\rho$. 
The denominator of Eq.~(\ref{eq:chemap}), 
$\langle \exp[F+i \theta] \rangle$, is given by 
\begin{eqnarray}
\left\langle e^F e^{i \theta} \right\rangle_{(T, \mu_q=0)}
= \frac{1}{{\cal Z}_{\rm GC}} \int dP \int dF \int d \theta \ 
e^F e^{i \theta} \bar{w}(P, F, \theta) ,
\label{eq:apdist}
\end{eqnarray}
where 
${\cal Z}_{\rm GC}=\int dP \int dF \int d \theta \ \bar{w}(P, F, \theta)$. 
Because we calculate this expectation value by the reweighting method 
using Eq.~(\ref{eq:evmultb}), the operator in the calculation of 
Eq.~(\ref{eq:apdist}) is a function of $P$, $F$ and $\theta$.

Since the partition function is real even at nonzero density, 
the distribution function is symmetric under 
the change from $\theta$ to $-\theta$.
Therefore, the distribution function is a function of $\theta^2$, 
e.g., $\bar{w}(\theta) \sim \exp[-(a_2 \theta^2 +a_4 \theta^4 
+a_6 \theta^6 + \cdots)].$
And, the distribution function is expected to be well approximated 
by a Gaussian function when the system size is 
sufficiently large in comparison to the correlation length. 
We assume the following distribution function in terms of
$\theta$ when $P$ and $F$ are fixed:
\begin{eqnarray}
\bar{w}(P, F, \theta) \approx \sqrt{\frac{a_2 (P, F)}{\pi}} 
\bar{w}'(P, F) \exp \left[-a_2 (P, F) \theta^{2} \right].
\label{eq:phap}
\end{eqnarray}
The coefficient $a_2 (P, F)$ is given by 
\begin{eqnarray}
\frac{1}{2a_2 (P', F')} &=&
\left. \int d \theta \ \theta^2 \ \bar{w}(P', F', \theta) \right/
\int d \theta \ \bar{w}(P', F', \theta)
\nonumber \\
&=&\frac{ \left\langle \theta^2 \delta (P'-P) 
\delta (F'-F) \right\rangle_{(T, \mu_q=0)} }{ \left\langle 
\delta (P'-P) \delta (F'-F) \right\rangle_{(T, \mu_q=0)}}
\equiv \left\langle \theta^2 \right\rangle_{(P, F)}.
\label{eq:a2}
\end{eqnarray}

The integration over $\theta$ can be carried out easily and we obtain 
the denominator of Eq.~(\ref{eq:chemap}), 
\begin{eqnarray}
\left\langle e^{F} e^{i \theta} \right\rangle_{(T, \mu_q=0)} 
&\approx& \frac{1}{{\cal Z}_{\rm GC}} \int dP \int dF \int d \theta \ 
\sqrt{\frac{a_2}{\pi}} \bar{w}'(P, F) e^{-a_2 \theta^2} 
e^F e^{i \theta}  
\nonumber \\
&=& \frac{1}{{\cal Z}_{\rm GC}} \int dP \int dF \ \bar{w}'(P, F) 
e^F e^{-1/(4a_2)}   
\nonumber \\
&=& \frac{1}{{\cal Z}_{\rm GC}} \int {\cal D}U 
e^{F} e^{-1/(4a_2(P, F))} 
(\det M(0))^{N_{\rm f}} e^{-S_g} \nonumber \\
&=& \left\langle e^{F} e^{-1/(4a_2(P, F))} 
\right\rangle_{(T, \mu_q=0)}.
\label{eq:gapp}
\end{eqnarray}
Since $\theta$ is roughly proportional to the size of the quark 
matrix $M$, the value of $1/a_2$ becomes larger as the 
volume increases. Therefore, the phase factor in this quantity decreases 
exponentially as a function of the volume. 
However, in this framework, the operator in Eq.~(\ref{eq:gapp}) is always real 
and positive for each configuration. This means that the expectation 
value is always larger than its statistical error.
Therefore, the sign problem is completely avoided if we can assume 
the Gaussian distribution of $\theta$. 
The effect from a non-Gaussian term $(a_4)$ is discussed in Appendix B 
for the case that $a_4/a_2$ is small. In this case, the effect from 
$a_4$ is at most $O(\mu_q^6)$, hence the non-Gaussian term may be 
neglected in the low density region even if $a_4$ is nonzero. 
The complex phase distribution of the quark determinant by chiral 
perturbation theory has been discussed in \cite{spli07}.
The Gaussian distribution is suggested in the 1-loop calculation.

The complex phase factor for the calculation of 
$\langle z_0 \exp[F+i \theta] \rangle$ in Eq.~(\ref{eq:chemap}) 
is calculated repeating the same procedure.
We consider the complex phase of $ z_0 \equiv |z_0| \exp [i \theta'] $, 
where $-\pi<  \theta' \le \pi$. Replacing $\theta$ with $\theta + \theta'$ 
in Eq.~(\ref{eq:gapp}), the suppression factor 
$\exp[-\langle (\theta + \theta')^2 \rangle /2]$ is estimated.

\begin{figure}[t]
\begin{center}
\includegraphics[width=3.1in]{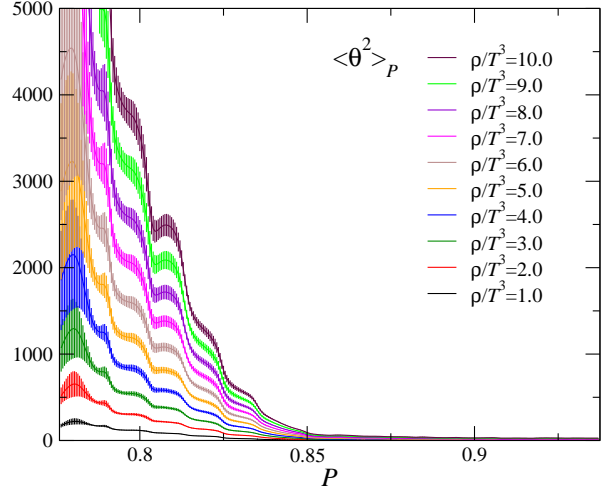}
\vskip -0.2cm
\caption{
Average of the square of the complex phase as a function of $P$ and $\rho/T^3$.
}
\label{fig:thhis}
\end{center}
\vskip -0.3cm
\end{figure}

We plot the result of the average of $\theta^2$ as a function of $P$ 
in Fig.~\ref{fig:thhis}, i.e. 
\begin{eqnarray}
\langle \theta^2 \rangle_{P'} \equiv \langle \theta^2 \delta (P-P') 
\rangle / \langle \delta (P-P') \rangle . 
\end{eqnarray}
We adopt the approximated delta function $\delta (P)$ used when we computed 
$\langle | z_0 | \rangle_P$ and $\langle F \rangle_P$ in Sec.~\ref{sec:analy}. 
Although we need to average $\theta^2$ as a function of $P$ and $F$ for 
the calculation of Eq.~(\ref{eq:gapp}), $F$ dependence is not considered 
in Fig.~\ref{fig:thhis}. 
It is found from this figure that $\langle \theta^2 \rangle_{P}$ 
decreases linearly as $P$ increases in the range of $P < 0.85$ 
and the phase fluctuation is small for $P > 0.85$. 
This means that the phase factor decreases exponentially as $P$ 
decreases for $P < 0.85$. 
This behavior is similar to that of $\langle F \rangle_{P}$ in 
Fig.~\ref{fig:ffhis}. 
Therefore, the weight factor $\exp [ F- \langle \theta^2 \rangle_{(P,F)}/2 ]$ 
suppresses the contribution from configurations having small $P$ 
for large $\rho/T^3$.

Moreover, this argument implies that configurations on which 
the sign problem is serious do not contribute to the actual 
calculations of expectation values. 
The reason is that the fluctuations of the complex phase 
$\langle \theta^2 \rangle_{(P,F)}$ are large on such configurations and 
the configurations are suppressed by the weight factor 
$\exp [- \langle \theta^2 \rangle_{(P,F)}/2 ]$.
Therefore, even if the error due to the Gaussian approximation of 
the complex phase distribution becomes visible when the phase fluctuations 
are large, the error does not affect to the practical calculations 
of expectation values so much.

Here, we should notice that the values of $P$ and $F$ are strongly correlated. 
We estimate the width of the distribution of $F$ for each $P$ and 
$\rho/T^3$ by calculating 
$\Delta F \equiv \sqrt{\langle (F- \langle F \rangle_P)^2 \rangle_P}$. 
Dashed lines above and below the solid line for $\langle F \rangle_P$ 
in Fig.~\ref{fig:ffhis} are 
the values of $\langle F \rangle_P \pm \Delta F$. 
Most of the configurations characterized by $P$ and $F$ are distributed 
in the narrow region between these two dashed lines.
Outside this bound an accurate calculation of 
$\langle \theta^2 \rangle_{(P,F)}$ is difficult, since the number of 
configurations is not enough for the average. 
However, if we consider that $F$ is approximately given as a function of 
$P$ on each configuration, $\langle \theta^2 \rangle_{(P,F(P))}$ will 
be a function of only $P$.

\section{Results and Discussions}
\label{sec:disc}

\begin{figure}[t]
\begin{center}
\includegraphics[width=3.1in]{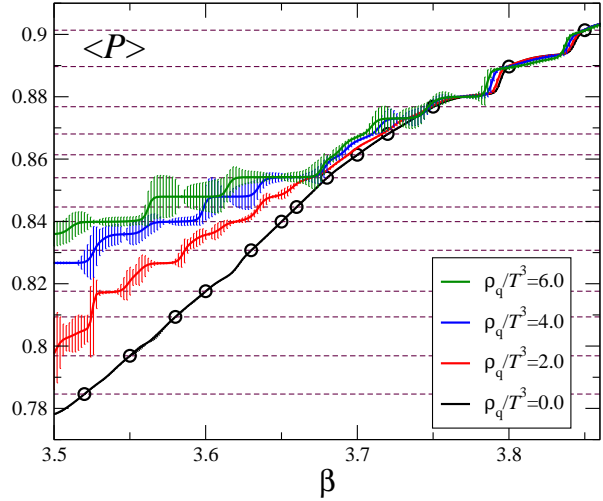}
\vskip -0.2cm
\caption{
Expectation value of the plaquette for each $\rho/T^3$.
Dashed lines are the peak positions of the plaquette distributions 
at $\mu_q=0$.
}
\label{fig:sg}
\end{center}
\vskip -0.3cm
\end{figure} 

\begin{figure}[t]
\begin{center}
\includegraphics[width=3.1in]{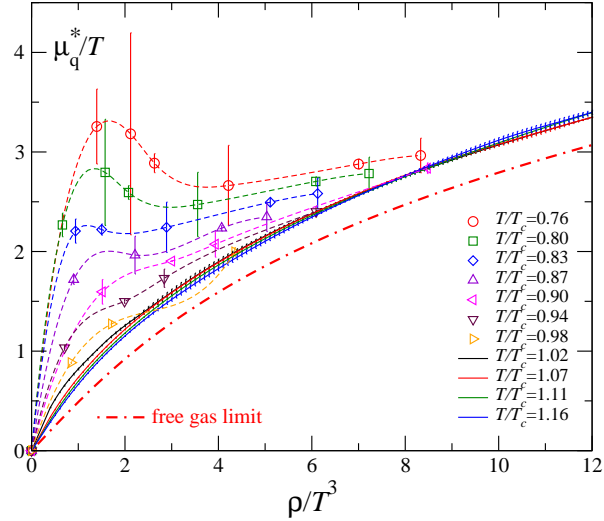}
\vskip -0.2cm
\caption{
Derivative of $\ln {\cal Z}_{C}$ as a function of the quark number density.
}
\label{fig:chepot}
\end{center}
\vskip -0.3cm
\end{figure}

We calculate the slope of $\ln {\cal Z}_{\rm C}$, i.e. $\mu_q^*/T$, 
using Eq.~(\ref{eq:chemap}). 
This quantity is given by the average of the saddle point $z_0$ multiplying 
the additional weight factor $\exp[F+i \theta]$.
The volume $V=16^3$ is sufficiently large, and we assume that the complex 
phase distribution of the reweighting factor is well approximated by 
a Gaussian function as discussed in Sec~\ref{sec:phase}. 
We then replace the weight factor by 
$\exp[F- \langle \theta^2 \rangle_{(F,P)}/2]$ to eliminate the sign problem. 
We find a saddle point $z_0$ numerically for each configuration, 
assuming $z_0$ exists near the real axis in the low density region of 
the complex $\mu_q/T$ plane. 

Before showing the result for $\mu_q^*/T$, it is worth discussing 
the effect of the weight factor using Eq.~(\ref{eq:sdm}). 
The weight factor can be approximately estimated by
\begin{eqnarray}
\langle \exp [ F - i \theta] \rangle_P w(P,\beta) \approx 
\Omega(P) \exp \left[ 6N_{\rm site} \beta P + \langle F \rangle_P 
- \langle \theta^2 \rangle_P/2 \right]
\end{eqnarray}
because 
$\ln \langle \exp [ F+i \theta ] \rangle_P \approx 
\langle F \rangle_P - \langle \theta^2 \rangle_P/2$ in the leading order and 
$w(P, \beta)$ can be written as $\Omega(P) \exp[6 \beta N_{\rm site} P]$ 
from Eq.~(\ref{eq:betarew}), where $\Omega(P)$ is the state density 
in terms of $P$.

Since the behavior of $\langle F \rangle_P - \langle \theta^2 \rangle_P/2$ 
for $P \simle 0.85$ and $P \simge 0.85$ is different, 
we consider these two regions separately. 
The configurations below and above $P \sim 0.85$ are generated in 
the low and high temperature phases, respectively.
In the region $P \simge 0.85$, the $P$ dependence of 
$\langle F \rangle_P - \langle \theta^2 \rangle_P/2$ is small. 
Therefore, the balance of the weight does not change. 
On the other hand, for $P \simle 0.85$, 
$\langle F \rangle_P - \langle \theta^2 \rangle_P/2$ increases linearly. 
This has the same effect as when $\beta$ changes to 
\begin{eqnarray}
\beta_{\rm eff} \equiv \beta + \left( \frac{d \langle F \rangle_P}{dP} 
- \frac{1}{2} \frac{d \langle \theta^2 \rangle_P}{dP} \right) 
\frac{1}{6N_{\rm site}}, 
\end{eqnarray}
and the derivatives of $ \langle F \rangle_P$ and 
$- \langle \theta^2 \rangle_P$ increase as $\rho$ increases. 
This means that the peak position of the probability distribution of $P$, 
shown in Fig.~\ref{fig:plhis} for $\mu_q=0$, moves to the right as $\rho$ 
increases, according to the change of the effective $\beta$.
This behavior is consistent with our usual expectation, i.e. 
a phase transition arises when the density is increased as well as 
with increasing temperature.

For the case that important configurations change, 
the multiparameter $(\beta)$ reweighting in 
Sec.~\ref{sec:analy} is effective, since the important 
configurations are automatically selected among all 
configurations generated at multi-$\beta$, and also this method is useful 
for the interpolation between the simulation points. 
We combine all data obtained at 16 points of $\beta$ using 
the multi-$\beta$ reweighting method. 

In order to observe the change of the important configurations, 
we calculate the expectation value of the plaquette with 
the additional weight factor in Fig.~\ref{fig:sg}, 
\begin{eqnarray}
\left\langle P \right\rangle (T, \rho)
\approx \frac{
\left\langle P \exp \left[ F + i \theta \right] \right\rangle_{(T, \mu_q=0)}}{
\left\langle \exp \left[ F + i \theta \right] \right\rangle_{(T, \mu_q=0)}}. 
\label{eq:plap}
\end{eqnarray}
The circles in Fig.~\ref{fig:sg} are $\langle P \rangle$ at $\rho/T^3=0$ 
computed without the multi-$\beta$ reweighting method. 
The solid lines which connect these circles are the interpolation 
using the multi-$\beta$ reweighting method. 
Of course, $\langle P \rangle$ for each simulation point and that 
obtained by the multi-$\beta$ reweighting method are consistent 
with each other at the simulation points. 
However, the line of the interpolation shows waves at some places 
between the circles. Such a wave appears where the configurations are 
missing in Fig.~\ref{fig:plhis}. 

For the calculation at finite $\rho/T^3$, we calculate the phase factor 
$e^{i \theta}$ from $\langle \theta^2 \rangle_P$ in Fig.~\ref{fig:thhis}. 
The $F$ dependence of $\langle \theta^2 \rangle_{(P,F)}$ is neglected 
because the values of $F$ is approximately given as a function of $P$ 
as seen in Fig.~\ref{fig:ffhis}. 
The results for $\langle P \rangle$ at $\rho/T^3=2.0, 4.0$ and $6.0$ are 
plotted from below. This quantity indicates the plaquette value of the 
most important configuration when the weight factor is modified. 

From this figure, we find that $\langle P \rangle$ becomes larger for 
$\beta < \beta_{pc} = 3.65$ and does not change very much for 
$\beta > 3.65$ when $\rho/T^3$ is increased. 
This means that, even when we measure an expectation value at small 
temperature (small $\beta$), the configurations generated by simulations 
at higher temperature (larger $P$) are used for the measurement 
at finite $\rho/T^3$. 
Moreover, for the measurement at sufficiently large $\rho/T^3$, the 
configurations generated in the low temperature phase are completely 
suppressed by the additional weight factor even when the temperature is small. 
This property explains why the phase transition happens when the density 
is increased while keeping the temperature fixed.

However, the errors on these results become larger as $\rho/T^3$ increases. 
One of the reasons may be the missing configurations between the peaks of 
the plaquette distributions.
As seen in Fig.~\ref{fig:plhis}, the configurations are not 
distributed uniformly in the range of $P$ which is necessary in 
this analysis, and correct results cannot be obtained if the important 
configurations are missing. 
At low temperature, the important value of $P$ changes very much as $\rho$ 
increases, therefore we calculate $\mu_q^*/T$ only when the expectation 
value of $P$ is at the peak positions of the plaquette distributions 
in Fig.~\ref{fig:plhis}. 
Dashed lines in Fig~\ref{fig:sg} are the peak positions.

Another important point is that $\langle |z_0| \rangle_P$ and 
$\langle F \rangle_P$ are strongly correlated with each other. 
Figure \ref{fig:rahis} is the average of $|z_0|$ as a function of 
the plaquette value of each configuration. 
$|z_0|$ increases as $P$ decreases. 
Because $D_1$ is purely imaginary, 
$-F ={\rm Re} [V(\bar{\rho}z_0 -N_{\rm f} N_t D_1 z_0)] +O(z_0^2)$  
becomes large as ${\rm Re} (z_0)$ increases, 
which is seen in Fig.~\ref{fig:ffhis}, and 
the contribution from the configurations which have large $z_0$ is 
suppressed by the additional weight factor. 
Although the value of $|z_0|$ for each configuration increases 
monotonically as a function of $\rho/T^3$, nontrivial behavior 
in $\mu_q^*/T$ is expected due to the suppression factor. 

We plot the result of $\mu_q^*/T$ in Fig.~\ref{fig:chepot} as a function 
of $\rho/T^3$ for each temperature $T/T_c (\beta)$. 
Dashed lines are cubic spline interpolations of these results.
The dot-dashed line is the value of the free quark-gluon gas in 
the continuum theory, 
\begin{eqnarray}
\frac{\rho}{T^3} = N_{\rm f} \left[ \frac{\mu_q}{T} 
+ \frac{1}{\pi^2} \left( \frac{\mu_q}{T} \right)^3 \right].
\end{eqnarray}
In this calculation, we neglected the $F$ dependence of 
$\langle \theta^2 \rangle_{(P,F)}$ because the values of $P$ and $F$ are 
strongly correlated.
The systematic error due to this approximation is discussed in Appendix C. 
The error seems to be small.

From Fig.~\ref{fig:chepot}, we find that a qualitative feature of $\mu_q^*/T$ 
changes around $T/T_c \sim 0.8$, i.e. $\mu_q^*/T$ increases monotonically 
as $\rho$ increases above 0.8, whereas it shows an s-shape below 0.8. 
This means that there is more than one value of $\rho/T^3$ for 
one value of $\mu_q^*/T$ below $T/T_c \sim 0.8$.
This is a signature of a first order phase transition. 
The critical point in the $(T, \mu_q)$ plane is estimated in \cite{eji07} 
by calculating the effective potential in terms of the plaquette value 
using the same configurations. 
The estimation is $(T/T_c, \mu_q/T) \approx (0.76, 2.5)$. 
Because the estimation from the effective potential is rather ambiguous, 
the difference between the new and old results of the critical temperature 
may be a systematic error. 
The critical value of $\mu_q^*/T$ is about $2.4$. 
This is almost consistent with the previous result by the different method.
The error from the truncation of the Taylor expansion of the quark 
determinant is discussed in \cite{eji07}. 
The difference between the results at $O(\mu_q^4)$ and $O(\mu_q^6)$ 
is found to be small at $\mu_q/T = 2.5$ for the data used in this study. 
Therefore, the error due to the truncation would not affect 
the qualitative conclusions. 
Although further studies including justifications of the 
approximations used in this analysis are necessary for more 
quantitative investigations, this result suggests the existence of 
the first order phase transition line in the $(T, \mu_q)$ plane.

\section{Conclusions}
\label{sec:conc}

We studied the canonical partition function as a function of $\rho/T^3$ 
performing an inverse Laplace transformation.
We analyzed the data obtained with two-flavors of p4-improved staggered 
quarks in \cite{BS05} and calculated the derivative of the canonical 
partition function with respect to $\rho$. 
The problems in this calculation were discussed. 
To avoid the problems, we adopted the following approximations. 
First, we estimate the quark determinant from the data of a Taylor 
expansion up to $O(\mu_q^6)$ because the direct calculation of the 
quark determinant is still difficult except on a small lattice. 
Although terms of higher than $\mu_q^6$ are omitted, this analysis is 
valid in the low density region. 
Second, we use a saddle point approximation for the inverse 
transformation, assuming the volume is sufficiently large. 
Third, we assume that the probability distribution of the complex 
phase of the operator in the calculation of $\mu_q^*/T^3$ can be 
well approximated by a Gaussian function. 

Using multiparameter reweighting method, we combined the configurations 
generated by $\mu_q=0$ simulations at 16 simulation points $(\beta)$ 
which cover a wide range of the temperature. 
It is found that the increase of $\mu_q^*/T^3$ becomes larger as 
the temperature decreases in the low density region. 
However, the contribution from the configurations generated at low 
temperature gradually decreases in the measurement of $\mu_q^*/T^3$ 
as $\rho/T^3$ increases even at low temperature. 
And, $\mu_q^*/T^3$ approaches the value of the free quark gas in the 
high density limit for all temperatures investigated in this study. 
The most interesting result is that $\mu_q^*/T^3$ as a function of 
$\rho/T^3$ shows an s-shape at $T \simle 0.8$. 
This means that the effective potential $V_{\rm eff}$ in terms of 
the density has two minima. 
Therefore, this result strongly suggests the existence of the first order 
phase transition line in the low temperature and high density region. 
Since the data we used in this study is obtained by a simulation with 
much heavier quark masses than the physical quark masses, simulations 
near the physical mass point are very important. 
It is also necessary to increase the accuracy of the approximations 
we have used in this study.

\section*{Acknowledgments}
I would like to thank 
F. Karsch, K. Kanaya, T. Hatsuda, S. Aoki, T. Izubuchi, 
and Y. Hidaka for discussions and comments.
This work has been authored under Contract No.~DE-AC02-98CH10886
with the U.S. Department of Energy.
I also wish to thank the Sumitomo Foundation for their 
financial assistance (No.~050408).

\section*{Appendix A: Multi-parameter $(\beta)$ reweighting method}

We discuss a method to combine configurations obtained by simulations 
with many $\beta$, following the method by Ferrenberg and Swendsen 
\cite{Swen89}.
We define a $\beta$-independent distribution function of the plaquette value,
\begin{eqnarray}
\Omega (P') = \int {\cal D}U \delta (P- P') (\det M(0))^{N_{\rm f}} .
\end{eqnarray}
The relation between $\Omega (P)$ and $w (P,\beta)$ in Eq.~(\ref{eq:pdis}) is 
$\Omega (P) \exp(6 \beta N_{\rm site} P) = w(P, \beta).$ 
Then, the expectation value of an operator ${\cal O}$ as a function 
of $P$ is given by the equation
\begin{eqnarray}
\left\langle {\cal O} \right\rangle_{\beta}
= \frac{1}{{\cal Z}_{\rm GC}}
\int {\cal O}(P) \Omega (P) e^{6 \beta N_{\rm site} P} dP 
\approx \frac{1}{N_{\rm conf}} \sum_{ \beta, \{ {\rm conf.} \}} {\cal O},
\hspace{5mm}
{\cal Z}_{\rm GC} = \int \Omega (P) e^{6 \beta N_{\rm site} P} dP, 
\label{eq:ev1beta}
\end{eqnarray}
where $N_{\rm conf.}$ is the number of configurations, and 
$\sum_{\beta, \{{\rm conf.}\}}$ denotes the sum of ${\cal O}$ over all 
configurations generated in a simulation at $\beta$.

The expectation value $\left\langle {\cal O} \right\rangle_{\beta}$ is also 
calculated from the data obtained by more than one simulation point, $\beta_i$
$(i=1,2, \cdots, N_{\beta})$. The Eq.~(\ref{eq:ev1beta}) is 
evaluated by 
\begin{eqnarray}
\int {\cal O}(P) \Omega (P) e^{6 \beta N_{\rm site} P} dP
&=& \sum_{i=1}^{N_{\beta}} \frac{N_i}{{\cal Z}_{\rm GC} (\beta_i)} 
\int {\cal O}(P) \Omega (P) 
\frac{e^{6 N_{\rm site} (\beta_i+\beta) P}}{
\sum_{j=1}^{N_{\beta}} N_j e^{6 N_{\rm site} \beta_j P} 
{\cal Z}_{\rm GC}^{-1} (\beta_j)} dP
\nonumber \\
& \approx & \sum_{i=1}^{N_{\beta}} 
\sum_{\beta_i, \{ {\rm conf.} \}} {\cal O}
\frac{e^{6 N_{\rm site} \beta P}}{
\sum_{j=1}^{N_{\beta}} N_j e^{6 N_{\rm site} \beta_j P} 
{\cal Z}_{\rm GC}^{-1} (\beta_j)} 
\end{eqnarray}
where $N_i$ is the number of configurations at simulation points $\beta_i$.
Hence, 
\begin{eqnarray}
\left\langle {\cal O} \right\rangle_{\beta}
\approx 
\frac{ \left\langle {\cal O} G(\beta,P) \right\rangle_{\rm all}
}{ \left\langle G(\beta,P) \right\rangle_{\rm all}},
\end{eqnarray}
Here, the weight factor $G(\beta, P)$ is
\begin{eqnarray}
G(\beta,P)=\frac{e^{6 N_{\rm site} \beta P}}{
\sum_{i=1}^{N_{\beta}} N_i e^{6 N_{\rm site} \beta_i P} 
{\cal Z}_{\rm GC}^{-1} (\beta_i)} ,
\end{eqnarray}
and $\left\langle \cdots \right\rangle_{\rm all}$ means the average over 
all configurations generated at all $\beta_i$.

The partition function ${\cal Z}_{\rm GC}(\beta_i)$ is 
determined by a consistency condition for each $i$,
\begin{eqnarray}
{\cal Z}_{\rm GC}(\beta_i) 
= \sum_{j=1}^{N_{\beta}} 
\sum_{\beta_j, {\rm \{ conf. \}}} G(\beta_i ,P) 
= \langle G(\beta_i, P) \rangle_{\rm all} .
\end{eqnarray}
This equation can be solved except for the normalization factor. 
We should note that $G(\beta, P)$ is independent of the simulation points 
$\beta_i$ at which the operators are measured, and important configurations 
for each calculation are selected by the weight factor automatically.

\section*{Appendix B: Effect from non-Gaussian terms in the phase factor}

We estimate the phase factor when the distribution is 
slightly different from Gaussian. 
We consider a distribution function with small $a_4(P,F)/a_2(P,F)$, 
\begin{eqnarray}
\bar{w}(P, F, \theta) \approx \sqrt{\frac{a_2}{\pi}} 
\left( 1-\frac{3a_4}{4a_2^2} +O \left[ \left( \frac{a_4}{a_2} \right)^2 
\right] \right)^{-1} \bar{w}'(P,F) 
e^{-(a_2 \theta^2 + a_4 \theta^4)}. 
\label{eq:wga}
\end{eqnarray}
In this case, the phase factor, $\exp[-1/(4a_2)]$, changes to 
\begin{eqnarray}
\int \bar{w}'(P,F)
\sqrt{\frac{a_2}{\pi}} \left( 1-\frac{3a_4}{4a_2^2} + \cdots \right)^{-1}
e^{i \theta} e^{-a_2 \theta^2 -a_4 \theta^4} d \theta 
\approx \bar{w}'(P,F) \exp \left( -\frac{1}{4a_2} 
+\frac{3a_4}{4a_2^3} -\frac{a_4}{16a_2^4} +O \left[ \left( 
\frac{a_4}{a_2} \right)^2 \right] \right) . 
\label{eq:a4parap}
\end{eqnarray}
and also the expectation value of $\theta^2$ for fixed $P$ and $F$ 
becomes 
\begin{eqnarray}
\langle \theta^2 \rangle_{(P,F)} = \frac{1}{2a_2} - \frac{3a_4}{2a_2^3} 
+O \left[ \left( \frac{a_4}{a_2} \right)^2 \right] .
\end{eqnarray}
From this equation, the term of $3a_4/(4a_2^3)$ in Eq.~(\ref{eq:a4parap}) 
is absorbed into $\langle \theta^2 \rangle /2$, 
hence the leading contribution from $a_4$ 
in the phase factor is $\exp[-a_4/(16a_2^4)]$. 
The value of $a_4$ can be evaluated by the Binder cumulant,
\begin{eqnarray}
B_4^{\theta} \equiv 
\frac{\left\langle \theta^4 \right\rangle_{(P,F)} }{ 
\left\langle \theta^2 \right\rangle_{(P,F)}^2 } 
=3- \frac{6a_4 (P,F)}{a_2^2 (P,F)} 
+O \left[ \left( \frac{a_4}{a_2} \right)^2 \right] .
\label{eq:a2a4}
\end{eqnarray}
Because $a_2^{-1} \sim \langle \theta^2 \rangle \sim O(\mu_q^2)$ for 
the chemical potential at the saddle point $z_0$, 
the effect from $a_4$ becomes larger as the density increases. 
Therefore, for the case of $a_4/a_2 \simle O(1)$ in the $\mu_q=0$ limit, 
$\exp[-a_4/(16a_2^4)]$ is $O(\mu_q^6)$ at most, i.e. 
\begin{eqnarray}
\left\langle e^{F} e^{i \theta} \right\rangle_{(T, \mu_q=0)} 
&=& \left\langle e^{F} \exp \left[ -\langle \theta^2 \rangle_{(P,F)}/2 
+O(\mu_q^6) \right] \right\rangle_{(T, \mu_q=0)}.
\end{eqnarray}
This argument suggests that the approximation by the Gaussian distribution 
is valid for the investigation of the low density region even if $a_4$ 
is nonzero, 
however the estimation of the range of $\rho$ in which the non-Gaussian 
contribution is small may be important as well as the application range 
of the Taylor expansion in $\mu_q$.

\section*{Appendix C: Suppression factor from complex phase fluctuation}

In the calculation of $\mu_q^*/T$, we need to calculate the suppression 
factor from the complex phase fluctuation, 
$\exp[ - \langle \theta^2 \rangle (P,F, \rho/T^3) ]$. 
This factor should be a function of $P, F$ and $\rho/T^3$, however 
we neglected the $F$ dependence in the calculation of Fig.~\ref{fig:chepot}. 
In this section, we discuss the error from this approximation. 
As shown in Fig.~\ref{fig:ffhis}, the values of $P$ and $F$ are strongly 
correlated. 
The solid line is the mean value of $F$ among the configurations having 
the plaquette value $P$ for each $\rho/T^3$. 
The dashed lines above and below the solid line show the fluctuations. 
Most of the configurations characterized by $P$ and $F$ are distributed 
in the narrow region between the two dashed lines. 
Therefore, we have approximated $\langle \theta^2 \rangle$ as a function 
of only $P$ and $\rho/T^3$ in Fig.~\ref{fig:thhis}. 

\begin{figure}[t]
\begin{center}
\includegraphics[width=3.1in]{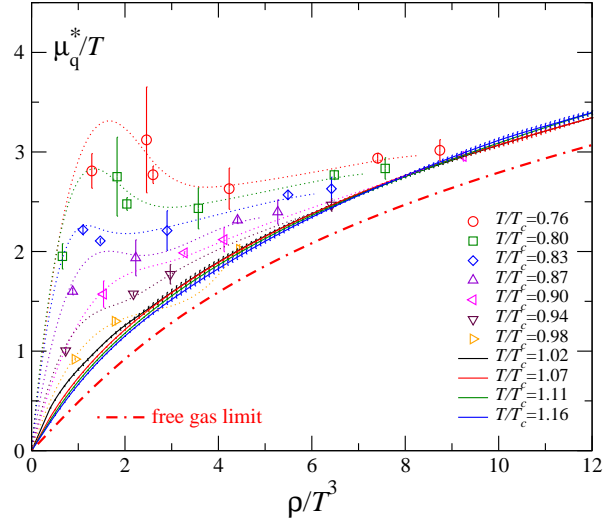}
\vskip -0.2cm
\caption{
Derivative of $\ln {\cal Z}_{C}$ as a function of the quark number density.
The estimation of $\langle \theta^2 \rangle (P,F)$ is different from 
that of Fig.~\ref{fig:chepot}. 
}
\label{fig:chepfd}
\end{center}
\vskip -0.3cm
\end{figure}

To estimate the importance of the $F$ dependence, we calculate $\mu_q^*/T$ 
using another approximation of $\langle \theta^2 \rangle$ which changes 
according to $F$ of each configuration. 
From the data of $\langle F \rangle_P$ as a function of $\rho/T^3$ for 
each $P$, which is shown in Fig.~\ref{fig:ffhis}, 
we find $\rho/T^3$ which gives $F$ for each configuration and 
find the value of $\langle \theta^2 \rangle$ at this $\rho/T^3$ using 
the data of Fig.~\ref{fig:thhis}. We then obtain $\langle \theta^2 \rangle$ 
as a function of $P$ and $F$, but the $\rho/T^3$ dependence is neglected. 
We calculate $\mu_q^*/T$ using this $\langle \theta^2 \rangle_{(P,F)}$ and 
compare the previous result to estimate the systematic error due to 
the approximation in $\langle \theta^2 \rangle$. 
The result is shown in Fig.~\ref{fig:chepfd}. 
The dotted lines are the spline interpolation in Fig.~\ref{fig:chepot}. 
The difference between the results of $\mu_q^*/T$ in Figs.~\ref{fig:chepot} 
and \ref{fig:chepfd} seems to be small. 
This suggests that the determination of $\langle \theta^2 \rangle$ 
with three parameter $(P, F, \rho/T^3)$ is not very important 
for the qualitative argument. 

Note: 
In the jackknife error estimation of this calculation, 
we have neglected the dispersion of $\langle \theta^2 \rangle_{(P,F)}$ 
among the jackknife ensemble. 
Therefore, statistical errors in Fig.~\ref{fig:chepfd} are smaller than 
those in Fig.~\ref{fig:chepot}. The small error does not mean that 
the analysis in this appendix gives better results with smaller 
statistical error. In fact, the errors in Fig.~\ref{fig:chepot} become 
the same size if we neglect the dispersion of $\langle \theta^2 \rangle$ 
in the jackknife analysis.

\end{document}